\documentclass[prb,twocolumn,showpacs,preprintnumbers,amsmath,amssymb]{revtex4}
\usepackage{graphicx}
\usepackage{dcolumn}
\usepackage{bm}

\begin{document}

\title{$T_{c}$ suppression and resistivity in cuprates with out of plane
defects}

\author{S. Graser}
 \affiliation{Department of Physics, University of Florida, Gainesville, FL 32611, USA}
\author{T. Dahm}
\affiliation{Institut f\"ur Theoretische Physik, Universit\"at T\"ubingen, T\"ubingen, Germany}
\affiliation{Institute for Solid State Physics, University of Tokyo, Kashiwanoha, Kashiwa, Chiba 277-8581, Japan}

\author{P. J. Hirschfeld}
\affiliation{Department of Physics, University of Florida, Gainesville, FL 32611, USA}

\author{L.-Y. Zhu}
\affiliation{Department of Physics, University of Florida, Gainesville, FL 32611, USA}

\date{\today}

\begin{abstract}
Recent experiments introducing controlled disorder into optimally
doped cuprate superconductors by both electron irradiation and
chemical substitution have found unusual behavior in the rate of
 suppression of the critical temperature $T_c$ vs. increase in
residual resistivity.  We show here that the unexpected discovery
that the rate of $T_c$ suppression vs. resistivity is stronger for
out-of-plane than for in-plane impurities may be explained by
consistent calculation of both $T_c$ and resistivity if the
potential scattering is assumed to be nearly forward in nature.
For realistic models of impurity potentials, we further show that
significant deviations from the universal Abrikosov-Gor'kov $T_c$
suppression behavior may be expected for out of plane impurities.

\end{abstract}


\maketitle

The destruction of superconductivity by disorder has been traditionally
used to probe the nature of the superconducting state. In classic
superconductors, pairbreaking is caused only by magnetic impurities \cite{Anderson},
and the functional form of the $T_c$ suppression, when plotted vs. impurity
concentration or change in normal state resistivity, is known to follow
the universal curve predicted by Abrikosov and Gor'kov (AG) \cite{AG}. In
unconventional superconductors, ordinary nonmagnetic impurities
are also expected to break pairs, and in the simplest approximation
where the impurities are treated as point-like (delta-function) potential
scatterers, the $T_c$ suppression also follows the AG form.

As in so many other respects, the experimental situation in the
cuprates agrees qualitatively with the simplest notions of what
should happen to $d$-wave superconductors in the presence of
disorder, but differs in some important details. For example, when
Zn is substituted for Cu in the Cu-O planes, $T_c$ is suppressed
rapidly as expected for a $d$-wave superconductor. Here,
``rapidly" means that the disorder-induced scattering rate
required to destroy superconductivity is on the order of the gap
scale rather than the Fermi energy $E_F$, as would be expected in
an $s$-wave system. Nevertheless, the initial slope of the $T_c$
vs. $\Delta \rho$ curve found in experiment is a factor of 2-3
smaller than the universal AG curve\cite{STolpygo:1996}. This
discrepancy has been attributed to scattering in higher angular
momentum channels by several authors, who modelled the scattering
potential with a separable form describing scattering in both $s$-
and a single higher $\ell$-wave channel \cite{Haran,Kulic97}. This
is a simple and tractable way of including the finite range of the
scatterers qualitatively, but is neither consistent by the
microscopics of screened impurities in the cuprates, nor capable
of treating the limit of extreme forward scattering, claimed to be
of relevance in the cuprate case \cite{Kee,Abrahams,Zhu}.
Furthermore, in these studies $T_c$ is calculated as a function of
the single-particle normal state scattering rate $1/\tau_N$, as
opposed to the transport rate $1/\tau_{tr}$ relevant for
comparison to resistivity measurements.

A further paradox was  reported recently by Fujita et al.
\cite{Fujita}, who showed that the rate of suppression
$dT_c/d\rho$ is significantly higher for out-of-plane cation
substituents in Bi$_2$Sr$_2$CuO$_6$ (Bi-2201) and (LSCO) than for
Zn in the same system. Since Zn is thought to be a near-unitary
scatterer in these materials, this is somewhat mysterious at first
sight. If one accounts for the fact that the out-of-plane defects
are poorly screened, and may act primarily as forward scatterers,
however, the increase of resistivity with defect concentration
will be slowed and one may be able to understand this discrepancy.
It is the primary purpose of this paper to correlate, within
simple models, the slower rates of both $T_c$ suppression and
resistivity increase in the case of near-forward scatterers to see
if light can be shed on this puzzle.

Other types of deviations from traditional AG behavior have been
observed and require explanation.  In the past several years,
electron irradiation studies have appeared which are able to
suppress superconductivity to zero in a controlled fashion in
contrast to early studies which studied only the initial slope of
the  $T_c$ suppression by disorder\cite{Giapintzakis}.
Rullier-Albenque et al.\cite{Rullier} used 2.5 MeV electron
irradiation to create defects throughout YBa$_2$Cu$_3$O$_7$
(YBCO-123) and YBa$_2$Cu$_3$O$_{6.6}$ samples. In neither sample
was an AG-type behavior for $T_c$ observed. In the optimally doped
sample, on the contrary, a remarkable linear behavior in $T_c$ vs.
$\rho$ was observed down to and including samples of vanishingly
small $T_c$. Because the nearly "normal" samples are expected to
have a small superfluid stiffness, this effect was interpreted
provisionally in terms of a phase fluctuation model proposed by
Emery and Kivelson \cite{Emery}. Deviations from AG theory at
small superfluid stiffness are to be expected even within the
framework of mean field theory, however, as pointed out by Franz
et al.\cite{Franzetal}, who studied the problem numerically and
included the self-consistent suppression of the order parameter
around each impurity site. These authors reported positive
curvature tails in $T_c$ vs. impurity concentration $n_i$ for
large $n_i$, in contrast to the negative curvature in the AG plot.

\section{Toy model}
To understand physically how some of  these effects might arise,
we consider a simple model of forward scattering by impurities in
which all effects can be calculated analytically.  We will assume
that in-plane impurities to be described by $\delta$-function like
isotropic potentials, and out-of-plane scatterers to be described
by an extended potential in the plane; this simply assumes that
the screened Coulomb potential created by out-of-plane impurities
produces a ``footprint" sensed by quasiparticles moving in the
CuO$_2$ planes.

As proposed by Kee \cite{Kee}, we interpolate between these cases
by considering a weak scattering potential $V_{\vec{k} \vec{k}'}$
which exists only at the Fermi surface $|\vec{k}|=k_F$, and cuts
off unless $\vec{k}$ and $\vec{k}'$ are sufficiently close
\begin{equation}
V(\phi,\phi')=\left\{ \begin{array}{cc}
v_{0} & \textrm{if}\,\,|\phi-\phi'|<\phi_{c}\\
0 & \textrm{otherwise}\end{array}\right.
\label{eq:pot_toymodel}
\end{equation}
The diagonal and off-diagonal self-energies in this model are then
\begin{eqnarray}
\Sigma_{0}(\phi,\omega_{n}) & = & -\Gamma\int_{\phi-\phi_{c}}^{\phi+\phi_{c}}d\phi'\frac{\tilde{i \omega}_{n}}{\sqrt{\tilde{\omega}_{n}^{2}+\tilde{\Delta}_{0}^{2}f(\phi')^{2}}} \label{eq:sigma0} \\
\Sigma_{1}(\phi,\omega_{n}) & = & \Gamma\int_{\phi-\phi_{c}}^{\phi+\phi_{c}}d\phi'\frac{\tilde{\Delta}_{0}f(\phi')}{\sqrt{\tilde{\omega}_{n}^{2}+\tilde{\Delta}_{0}^{2}f(\phi')^{2}}} \label{eq:sigma1}
\end{eqnarray}
where $\Gamma = \pi n_i N_0 v_0^2$, with 
$N_0$ the density of states at the Fermi level. In
Eqs.~(\ref{eq:sigma0}) and (\ref{eq:sigma1}), the renormalized
Matsubara frequencies and gap magnitudes are
$i \tilde{\omega}_n=i \omega_n- \Sigma_0$  and $\tilde{\Delta}_0 =
\Delta_0 + \Sigma_1$, and we assume a $d$-wave form of the
unrenormalized order parameter, $f(\phi) = \cos 2\phi$.  Note we
work in units where Boltzmann's constant $k_B=1$. The critical
temperature is determined as usual from the linearized gap
equation
\begin{equation}
\Delta_{0} = V_{d}N_{0}2\pi T_{c}\sum_{\omega_{n}>0}^{\omega_{c}}
\frac{1}{2\pi}\int d\phi'f(\phi')^{2}\frac{\tilde{\Delta}_{0}}{\tilde{\omega}_{n}}
\label{eq:gap_equation}
\end{equation}
where the $d$-wave pairing interaction is given by
$\left(V_{d}N_{0}\right)^{-1} = 1/2 \ln(2e^{\gamma}\omega_{c}/\pi T_{c0})$,
and $T_{c0}$ is $T_c$ in the absence of impurities. If we now take $\phi_c \ll 1$,
and $T_c \lesssim T_{c0}$ we find to leading order
$\tilde{\Delta}_0/\tilde{\omega}_n = \Delta_0/\omega_n$,
i.e. there is no pairbreaking by pure forward scattering, analogous to Anderson's
theorem in an $s$-wave superconductor with isotropic nonmagnetic impurities.

Since we are only interested in calculating the critical temperature we can neglect
quadratic variations of the order parameter and therefore perform the angular integrations
in the definition of the self-energies. After solving the resulting set of equations for the ratio
$\tilde{\Delta}_0/\tilde{\omega}_n$ that enters Eq.~(\ref{eq:gap_equation}), we can write
\begin{equation}
1 =  V_d N_0\pi T_c\sum_{\omega_n>0}^{\omega_c} \frac{1}{\omega_n
+ 2 \Gamma \phi_c - \Gamma \sin(2 \phi_c)}
\label{eq:gap_eq_expanded}
\end{equation}
It is worthwhile noting at this point that an expansion in powers
of $\phi_c$ applied to (\ref{eq:gap_eq_expanded}) as performed in
Ref. \onlinecite{Kee} is appropriate except when $T_c\rightarrow
0$. Following this approach would lead to the incorrect conclusion
that the $T_c$ suppression is linear all the way to $T_c=0$ in the
forward scattering $\phi_c\rightarrow 0$ limit.  Instead, it is
clear that the expression (\ref{eq:gap_eq_expanded}) may be summed
exactly, leading to  a modified AG result of the form
\begin{equation}
\ln \frac{T_c}{T_{c0}} = \psi \left( \frac{1}{2} \right) - \psi
\left( \frac{1}{2} +\frac{\Gamma (\phi_c - \sin \phi_c \cos
\phi_c)}{\pi T_c} \right)\label{eq:modifiedAG}
\end{equation}
where the pair breaking parameter in the forward scattering limit
($\phi_c \ll 1$) is of the order of $\phi_c^3 $:
\begin{equation}
\frac{\Gamma (\phi_c - \sin \phi_c \cos \phi_c)}{\pi T_c}
\approx \frac{2 \Gamma \phi_c^3 }{3 \pi T_c}
\end{equation}
Thus we see that the effective pairbreaking rate will decrease
dramatically as the scattering becomes more forward in nature.
This is indeed consistent with the conclusions of earlier
works\cite{Haran}, which concluded that anisotropic scattering
slows $T_c$ suppression for  fixed impurity concentration.

\begin{figure}
\includegraphics[width=1.0\columnwidth]{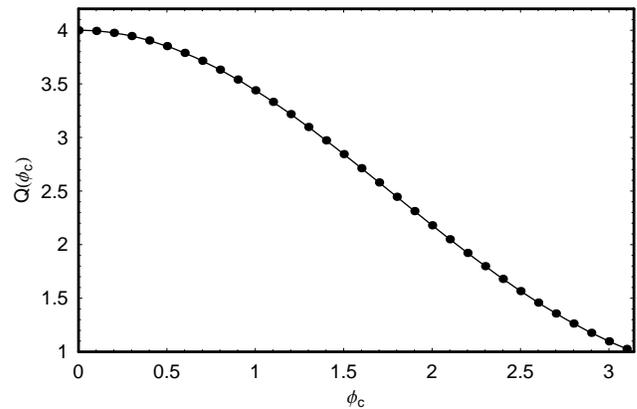}
\caption{\label{fig:Q} The factor $Q(\phi_c)$ as a function of the
maximum scattering angle $\phi_c$. It reflects the decrease of the
initial slope of $T_c$ suppression as a function of resistivity if
one approaches the forward scattering limit ($\phi_c \rightarrow
0$).}
\end{figure}

It remains, however, to calculate the dependence of the normal
state resistivity on the impurity scattering within the same
framework if one wishes to compare directly to experiments where
pairbreaking is measured by the increase in residual resistivity.
For an isotropic 2D Fermi surface this may be written as
\begin{equation}
\rho  =  \frac{2m^{2}}{e^{2}p_{F}^{2}}\int d \phi \; d\phi'(1-\cos(\phi'-\phi))n_{i}|V(\phi,\phi')|^{2}
\label{eq:rho_phi_phip}
\end{equation}
leading us to the expression
\begin{equation}
\rho  = \frac{\hbar}{e^{2}}\frac{2\Gamma}{E_{F}}\left( 2 \phi_c - 2 \sin \phi_c \right)
\end{equation}
so that $T_c$ may be expressed directly in terms of the impurity-induced
resistivity as
\begin{equation}
\ln \frac{T_c}{T_{c0}} = \psi \left( \frac{1}{2} \right)
- \psi \left( \frac{1}{2} + Q(\phi_c) \rho \frac{e^2 E_F}{4 \pi \hbar T_c} \right)
\label{eq:tc_vs_rho_forward}
\end{equation}
where the factor $Q(\phi_c)$ is given as
\begin{equation}
Q(\phi_c) = (\phi_c - \sin \phi_c \cos \phi_c)/(\phi_c - \sin \phi_c)
\end{equation}
It ranges from $Q(0)=4$ in the forward scattering limit to
$Q(\pi)=1$ in the case of isotropic scattering and its dependence
on $\phi_c$ is shown in Fig.~\ref{fig:Q}.
 The factor $Q(\phi_c)$
is also directly related to the initial slope of the $T_c$
suppression that can be derived from
Eq.~(\ref{eq:tc_vs_rho_forward}) as
\begin{equation}
\frac{T_{c}}{T_{c0}} \approx 1-\frac{\pi}{8} Q(\phi_c)
\rho\left(\frac{E_{F}}{T_{c0}}\frac{e^{2}}{\hbar}\right)
\label{eq:TcSupp}
\end{equation}

\begin{figure}
\includegraphics[width=1.0\columnwidth]{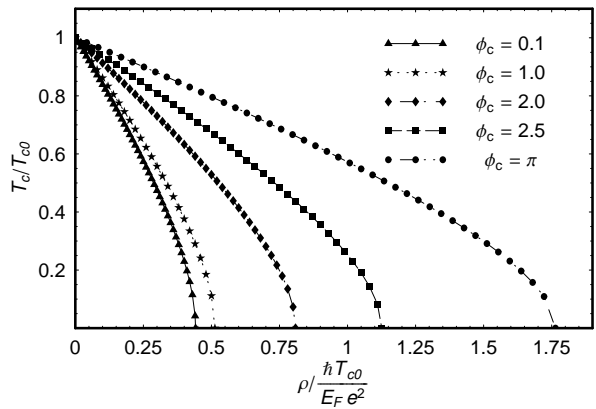}
\caption{\label{fig:toy_model} $T_c$ suppression as a function of
resistivity calculated within the toy model. We find a decrease
of the initial slope of the $T_c$ suppression as a function of resistivity
if we approach the forward scattering limit.}
\end{figure}


We see that although the rate of $T_c$ suppression (or resistivity
increase) is certainly much slower for forward scatterers than in
the isotropic scattering case, the dependence on the resistivity
is stronger for the former class of defects -- within our model
about a factor of four. This may indeed explain the results of
Fujita et al. \cite{Fujita}, since in these experiments out of
plane (more forward scattering) and in-plane (isotropic)
impurities are studied in separate samples.
In Fig.\ref{fig:toy_model} we show the suppression of $T_c$ as a
function of resistivity for different values of $\phi_c$ as given
by Eq.~(\ref{eq:tc_vs_rho_forward}).  While the initial slope is
indeed seen to increase as the scattering is made more
anisotropic, approaching the value of 4 as $\phi_c\rightarrow 0$,
we see also that within the toy model there is no deviation from
the {\it form} of the AG curve, as exhibited explicitly in Eq.
(\ref{eq:modifiedAG}).

\section{Realistic model}
To confirm the above intuition, and get a sense of
the range of behavior possible in real systems, we
now consider a more realistic model consisting of
randomly distributed out-of-plane impurities
with Yukawa potentials $V_i = V_0 \exp(- \kappa r_i)/r_i$,
where $r_i$ is the distance from a dopant atom to the lattice site
$i$ in the plane. $1/\kappa$ gives the screening length of the
impurity potential and the forward scattering case
($\kappa \rightarrow 0$) as well as the
isotropic scattering case ($\kappa \rightarrow \infty$) are
included with this specific choice of the
impurity potential. The Fourier components
of the screened Coulomb potential can be written as
\begin{equation}
\left|V(\vec{k},\vec{k}')\right|^{2}=
\frac{\left|V_{0}\right|^{2}}{\left|\vec{k}-\vec{k}'\right|^{2}+\kappa^{2}}
\label{Vrealistic}
\end{equation}

\begin{figure}
\includegraphics[width=1.0\columnwidth]{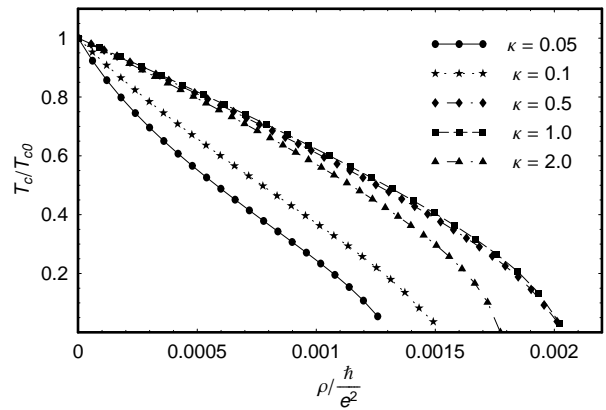}
\caption{\label{fig:realistic_model} $T_c$ suppression as a
function of resistivity calculated within a tight binding model.
$1/\kappa$ gives the range of the scattering potential, where
$\kappa \ll 1$ corresponds to the forward scattering and $\kappa
\gg 1$ to the isotropic scattering case. The non-monotonic
behavior of the curves for different $\kappa$ results from the
competing effects of a reduction of the downward curvature with a
simultaneous decrease of the initial slope for decreasing
$\kappa$. The critical temperature of the pure sample is chosen to
be $T_{c0}=0.01 t$.}
\end{figure}

We use a square lattice to mimic the copper-oxide plane and
include up to next-nearest hopping terms in numerical
computations. The  electronic band is taken to be
$\epsilon_{k}=-2t\left(\cos k_{x}+\cos k_{y}\right)- 4t'\cos
k_{x}\cos k_{y}-\mu$, with $t'=-0.3\, t$ and $\mu=-1.1\, t$ in
order to mimic the known shape of hole-doped cuprate Fermi
surfaces near optimal doping. The $T_c$ suppression is calculated
by solving the gap equation self-consistently:
\begin{equation}
\Delta_{\vec{k}}=\frac{1}{\Lambda\beta}\sum_{\omega_n,\vec{k}'}
V_{\vec{k},\vec{k}'}^d\frac{\tilde{\Delta}_{k'}}{\tilde{\omega}_n^2+\epsilon_{k'}^2}
\end{equation}
with an effective electron-electron pairing interaction
$V_{\vec{k},\vec{k}'}^{d}=V^{d}f(\vec{k})f(\vec{k}')$. The choice of the
symmetry function $f(\vec{k})=1/2 \left(\cos k_{x}-\cos k_{y}\right)$
leads to a $d_{x^2-y^2}$ symmetry of the order parameter. The diagonal and
off-diagonal parts of the self-energy are given by
\begin{eqnarray}
\Sigma_0 (\vec{k},\omega) & = & n_i \sum_{\vec{k}'}\left|V(\vec{k},\vec{k}')\right|^2
\frac{\tilde{\omega}}{\tilde{\omega}^2-\epsilon_{k'}^{2}-\tilde{\Delta}_{k'}^2}\\
\Sigma_1 (\vec{k},\omega) & = & n_i \sum_{\vec{k}'}\left|V(\vec{k},\vec{k}')\right|^2
\frac{\tilde{\Delta}_{k'}}{\tilde{\omega}^2-\epsilon_{k'}^2-\tilde{\Delta}_{k'}^2}
\end{eqnarray}
where we can neglect quadratic variations of $\tilde{\Delta}_k$
since we are only interested in the region near $T_c$. To compare
the suppression of $T_c$ to the increase of the normal conducting
resistivity we have to calculate the resistance within the same
model. For the transport rate $\tau^{-1}(\vec{k})$ we use an
approximation introduced by J. M. Ziman \cite{Ziman} that expands
the quasiparticle scattering rate by an additional scattering-in
term,
\begin{equation}
\frac{1}{\tau_{k}}=\frac{2\pi}{\hbar}
n_{i}\int\frac{dk_x'dk_y'}{(2\pi/a)^2}\delta\left(\epsilon_{k'}\right)
\left(1-\frac{\vec{v}_F(\vec{k})\cdot\vec{v}_F(\vec{k}')}{\left|\vec{v}_F(\vec{k})\right|
\left|\vec{v}_F(\vec{k}')\right|}\right) \left|V_{kk'}\right|^2
\end{equation}
and that has proven to be very accurate even for highly anisotropic
transport rates\cite{Lawrence}. In the expression for the
transport rate we have replaced the full $T$ matrix by the single impurity
scattering potential $V_{kk'}$. The conductivity can then be
calculated from
\begin{equation}
\rho^{-1} = \sigma_{xx}=e^2\int\frac{d^2 k}{(2\pi)^2}\tau_{k} v_{F,x}^2 \delta\left(\epsilon_k \right)
\end{equation}
assuming a cubic symmetry of the transport tensor.

\begin{figure}
\includegraphics[width=1.0\columnwidth]{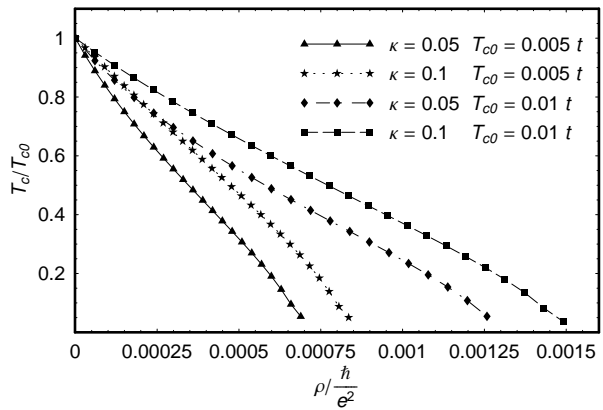}
\caption{\label{fig:realistic_model_Tc0compare} $T_c$ suppression
as a function of resistivity in the forward scattering limit for
two different values of $T_{c0}$. The upward curvature that is
visible especially for very low values of $\kappa$ diminishes for
lower values of $T_{c0}$ and the $T_c$ suppression as a function
of resistivity becomes more linear.}
\end{figure}

In Fig.~\ref{fig:realistic_model}, the suppression of $T_c$ as a
function of resistivity is shown for the tight binding model
discussed in this section and $T_{c0}=0.01 t$. As we have already
seen for the toy model, the initial rate of $T_c$ suppression vs.
resistivity increases with  increasing range of the impurity
potential $1/\kappa$, and seems to be a robust feature of the
$T_c$ suppression due to forward scattering processes. However the
particular shape of the $T_c$ suppression curve depends on the
details of the considered model, e.g. the band structure, the
doping strength or the functional form of the scattering
potential. In Fig.\ref{fig:realistic_model_Tc0compare}, it is
shown that by approaching the weak coupling limit  $T_{c0}\ll E_F$
due to a lowering of $T_{c0}$, we find a more linear dependence of
the critical temperature on the resistivity, reminiscent of the
linearity found in the electron irradiation experiments. At
present, we do not have an analytical understanding of the origins
of this quasilinearity.

\section{Conclusions}

We have investigated the disorder-induced reduction of the
critical temperature in high $T_c$ compounds due to out-of-plane
disorder, assuming that the out-of-plane defects act primarily as
elastic forward scatterers.  These calculations may be important
not only to understand experiments where this type of disorder is
varied systematically, but also to understand $T_c$'s in systems
which are intrinsically disordered by the doping process.
Calculating both the normal state resistivity as well as the $T_c$
suppression in both a simple toy model and for more realistic
scattering potentials and bands, we found that although the effect
of forward scattering on both quantities is smaller than for
isotropic scattering, the $T_c$ suppression as a function of
resistivity is stronger for forward than for isotropic scatterers.
In the case of the toy model, we showed that the suppression is
AG-like, with modified pairbreaking parameter.  The basic physics
of the more rapid initial suppression of $T_c$ vs $\rho$ was
confirmed numerically using more realistic Yukawa-type impurity
potentials within a tight binding model.

Our result stands in apparent contradiction to earlier work
comparing $T_c$ vs $\rho$ for anisotropic potentials
scatterers\cite{Kulic97}, where only $s$ and $d$-wave components
of the scattering potential were retained. These authors concluded
that the greater the ratio of potentials $V_d/V_s$, the weaker
would be the suppression of $T_c$ vs $\rho$. While this model is
mathematically consistent, it is physically unrealistic and does
not yield generic results.  As shown in the Appendix, more
realistic potentials may be expanded in Fermi surface harmonics,
but to obtain the proper (weaker) enhancement of the resistivity
with disorder it is important to retain the $p$-wave component as
well.  If this is done for a generic potential, the result of this
paper obtains.

\begin{figure}
\includegraphics[width=1.0\columnwidth]{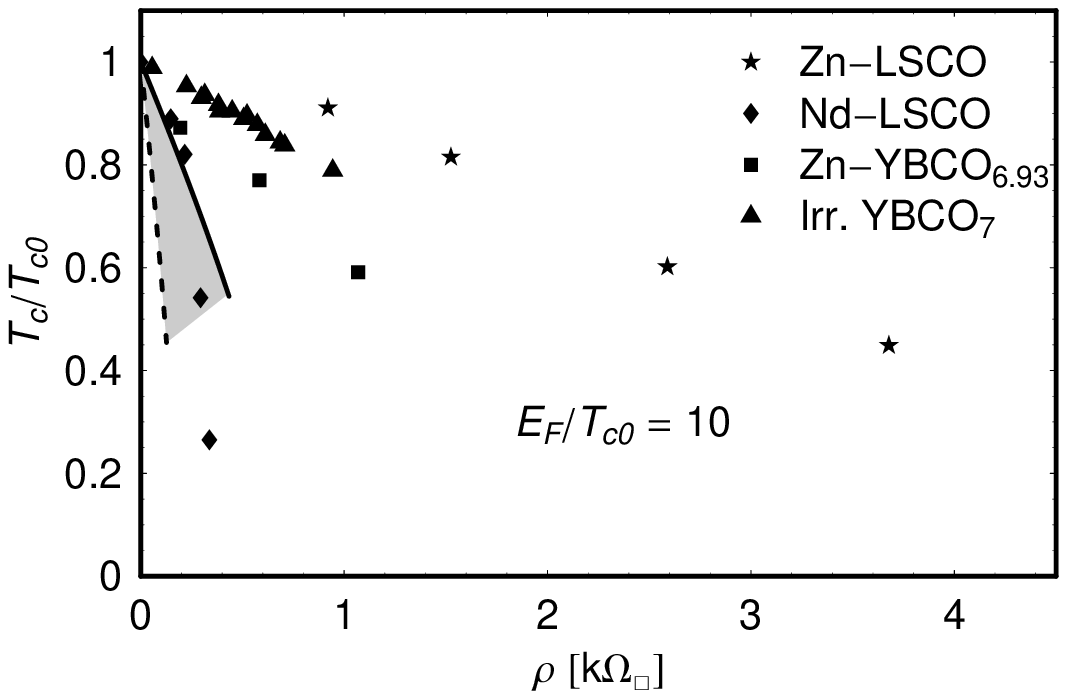}
\includegraphics[width=1.0\columnwidth]{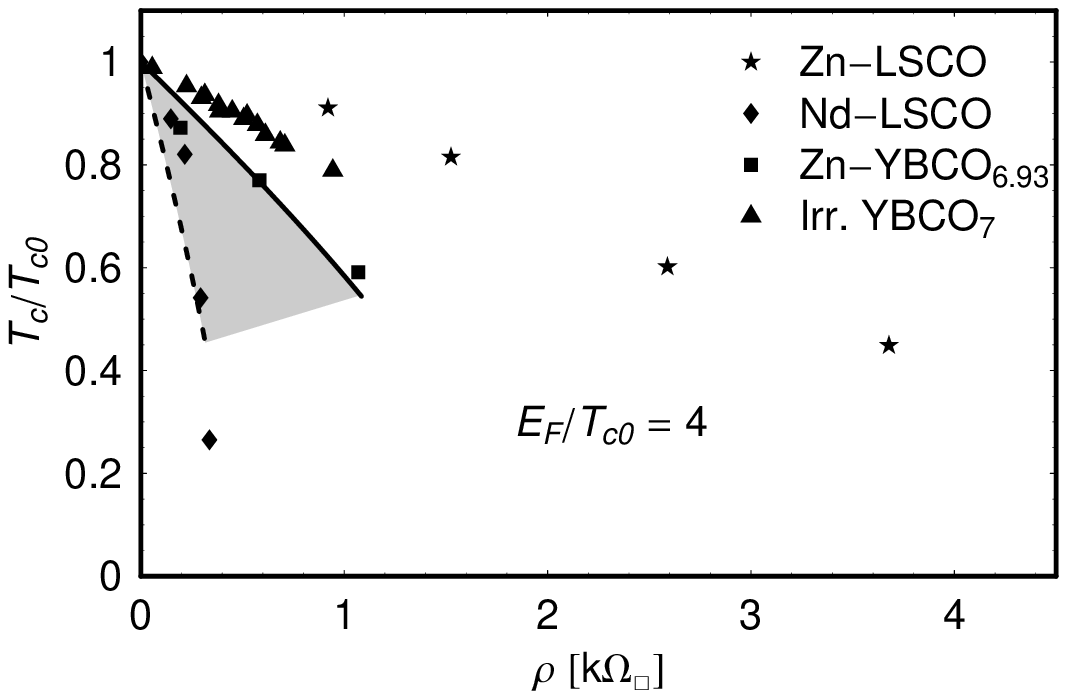}
\caption{\label{fig:experiment_compare} Comparison of experimental
data on $T_c$ suppression in optimally doped cuprates.  Diamonds:
Nd-LSCO \cite{Fujita}; squares: Zn-YBCO\cite{Fukuzumi}; triangles:
electron-irradiated YBCO \cite{Rullier}; stars: Zn-LSCO
\cite{Fujita}. Solid lines: toy model  result for isotropic
scatterers; dashed lines: for purely forward scatterers.  Top
(bottom) panel, $E_F/T_{c0} = 10 (4).$ }
\end{figure}

This point leads us to an important conclusion regarding $T_c$
suppression experiments making use of  Zn to replace Cu.  It has
been recognized for some time that the $T_c$ suppression rate vs.
residual resistivity for Zn is  smaller than predicted by AG
theory for pointlike scatterers in a $d$-wave
superconductor\cite{STolpygo:1996} by about a factor of three in
optimally doped YBCO, and even smaller in other materials like
LSCO.  We have now argued that the "conventional" explanation,
based on highly anisotropic scattering by Zn impurities, is
unlikely to be correct.  It seems, therefore, that even in the
optimally doped cuprates the deviation must be ascribed to effects
of strong correlations or strong coupling superconductivity.  
Within Eliashberg theory,\cite{Schachinger}  the AG pair-breaking
parameter is renormalized $\Gamma \rightarrow \Gamma/(1+\lambda)$, 
where $\lambda$ is the dimensionless coupling, 
while the disorder-induced resistivity
change is unrenormalized to leading order. The physical origin
of this effect is not clear in the cuprate context, however,
Kulic and Oudovenko\cite{Kulic97} argue that renormalizations of the
scattering vertex due to strong interactions within the t -J
model introduce significant suppression of the transport rate
induced by impurities. While this cannot explain the effects
of forward potential scattering for out of plane scatterers as
discussed here, since the renormalization in their model is
independent of the anisotropy of the scattering, it may be
part of the solution to the Zn problem. Another perspective
on the same physics to explain this effect may involve the
low-energy spin fluctuations known to be induced by disorder
in strongly correlated systems \cite{HAlloul:2007}. 
No theoretical work which
considers both $T_c$ and residual $\rho$ is available at this
writing, however; it is difficult to predict {\it a priori} which
quantity will be more strongly influenced by correlations.

We summarize the situation comparing the current theory to
experiments in Fig. \ref{fig:experiment_compare}.  Since we are
primarily interested in order-of-magnitude physics, we discuss
only the initial slopes of $T_c$ suppression measured in various
experiments, and compare to the toy model result Eq.
(\ref{eq:TcSupp}).  To obtain a fit one must make an assumption
about the parameter $E_F/T_{c0}$ which enters this expression.  A
reasonable choice for the high-$T_c$ materials is $E_F/T_{c0}=10$,
and this choice puts most of the data on Zn-YBCO about a factor of
3 higher in slope, as found by previous authors.  The range of
$T_c$ suppression initial slopes within the current approach is
then  shown in the Figure in gray, ranging from the isotropic
result to the extreme forward scattering result.  For comparison,
a set of curves is also given for the unphysical case
$E_F/T_{c0}=4$.  While these parameters should not be taken too
seriously in a quantitative sense, it seems clear that a) a
hitherto unaccounted for physical mechanism is required to explain
the data on Zn-substituted samples; and b) the effect of
out-of-plane scatterers within the present model can only account
for a factor of 4 or so increase in the magnitude of the slope.
Thus our work has explained qualitatively the
paradoxical result that the out of plane scatterers generically
reduce $T_c$ more quickly than in-plane relative to residual
resistivity, there is still a  quantitative question remaining
regarding the magnitude of the suppression in both cases.

The final experimental result we have attempted to discuss is the
fascinating linear $T_c$ vs $\rho$ suppression measured on 
electron-irradiated single crystals of optimally doped YBCO by 
Rullier-Albenque et al\cite{Rullier}. The deviation
from the AG pairbreaking result was attributed by these authors to
phase fluctuations according to a model put forward in Ref.
\onlinecite{Emery}. This is plausible for the underdoped sample
where phase fluctuations are expected to be strong, and for the
highly irradiated optimally doped sample, where the superfluid
density is also small. We note, however, that within this scenario
it is puzzling and must be regarded as accidental that the initial
$T_c$ slope in the optimally doped sample, due to pairbreaking, is
the same as the final slope before superconductivity disappears.
It is therefore equally likely, in our opinion, that the
quasilinearity found by these authors is due to the effects of
out-of-plane defects created by the electron irradiation, as
discussed here, and/or the effects of order parameter suppression
as discussed in Ref. \onlinecite{Franzetal}. ~ \vskip 1cm ~

\begin{acknowledgments}
The authors acknowledge valuable conversations with H.
Alloul, J. P. Carbotte, H. Eisaki, M. Kulic, E. Nicol, and F.
Rullier-Albenque. Partial support for this research was provided
by DOE Grant No. DE-FG02-05ER46236.
\end{acknowledgments}

\appendix*

\section{Fourier expansion of the impurity potential}

In previous works the effect of forward scattering on the critical
temperature has been studied by including higher angular momenta
of the impurity potential when calculating the self-energies. In
this section we show the link between the results of our toy model
and an approach, where the impurity potential is decomposed in its
relevant Fourier components. Particularly we want to point out the
importance of the $p$-wave component of the impurity potential for
calculating the normal conducting resistivity that has been
neglected in previous works and that is the key in understanding
the stronger suppression of the critical temperature as a function
of resistivity in the case where forward scattering is the
dominant scattering process.

Starting with the square of the impurity potential that is given
in Eq.~(\ref{eq:pot_toymodel}) we can expand it in a Fourier
series of its two arguments $\phi$ and $\phi'$
\begin{eqnarray}
|V(\phi,\phi')|^2 & = & v_0^2 V_0 +
v_0^2 \sum_{k=1}^{\infty}  V_k \\
& \times & \left[\cos (k \phi) \cos (k \phi') + \sin (k \phi) \sin
(k \phi') \right] \nonumber
\end{eqnarray}
with $V_0 = \phi_c/\pi$ and $V_k = 2 \sin(k \phi_c)/k \pi$. Using
this expansion to calculate the self-energies near $T_c$ we notice
that the integration projects out only the $s$- and the $d$-wave
part of the impurity potential, leading to
\begin{equation}
\Sigma_0 = 2 \pi \Gamma V_0
\end{equation}
and
\begin{equation}
\Sigma_1 (\phi,\omega_n) = \pi \Gamma V_2 \frac{\tilde{\Delta}_0
\cos(2 \phi)}{\tilde{\omega}_n}
\end{equation}
Solving for $\tilde{\Delta}_0/\tilde{\omega}_n$ and performing the
Matsubara frequency summation in the gap equation we find
\begin{equation}
\ln \frac{T_c}{T_{c0}} = \psi \left( \frac{1}{2} \right) - \psi
\left( \frac{1}{2} +\frac{\Gamma (2 V_0 - V_2)}{2 T_c} \right)
\end{equation}
This result is in agreement with the results of Kulic and
Oudovenko and their parameters $\Gamma_s$ and $\Gamma_d$ can now
be directly compared to the toy model parameter $\phi_c$:
\begin{equation}
\Gamma_s = 4 \Gamma \phi_c, \; \Gamma_d = 2 \Gamma \sin(2 \phi_c)
\end{equation}
To calculate the normal conducting resistivity one has to weight
the quasiparticle scattering probability by its impact on the
resistivity leading to Eq.~(\ref{eq:rho_phi_phip}). It is obvious
that in the case of pure forward scattering the increase of
resistivity with impurity concentration is much slower than for
isotropic scattering, a fact that has been taken into account by
the factor $1-\cos(\phi - \phi')$ that projects not only the
$s$-wave part but also the $p$-wave part of the impurity potential
out of the Fourier expansion. The resistivity can then be written
as
\begin{equation}
\rho = \frac{4 \pi \hbar}{e^2 E_F} \Gamma \left(V_0 - \frac{1}{2}
V_1 \right)
\end{equation}
Solving this equation for $\Gamma$ and inserting it in the
expression for the $T_c$ suppression leads to
Eq.~(\ref{eq:TcSupp}) where $Q(\phi_c)$ can be expressed by the
$s$-, $p$- and $d$-wave part of the impurity potential as
$Q(\phi_c)=(2 V_0 - V_2)/(2 V_0 - V_1)$. The dependence of
$Q(\phi_c)$ on the maximum scattering angle $\phi_c$ is shown in
Fig.~\ref{fig:Q}. It shows that the effect of impurity scattering
on the resistivity is drastically reduced in the forward
scattering limit due to the $p$-wave character of the
scattering-in term while the effect on the $T_c$ reduction is not
as strong, since the   only anisotropic component which plays a
role in the latter case is $d$-wave.

\end{document}